\documentclass[11pt]{amsart}
\usepackage{geometry}                
\usepackage{graphicx}
\usepackage{amssymb}
\usepackage{epstopdf}
\makeatletter
\renewcommand{\@biblabel}[1]{\quad#1.}
\makeatother

\date{}

\usepackage{lastpage,fancyhdr,graphicx}
\usepackage{epstopdf}
\usepackage{changepage}
\usepackage{graphicx,color}
\usepackage{url,geometry}
\usepackage{framed}
\usepackage{color}
\usepackage{appendix}
\usepackage{threeparttable}

\newcommand{\I}{\mathcal{I}} 
\newcommand{\J}{\mathcal{J}} 
\newcommand{\M}{\mathcal{M}} 
\newcommand{\m}{\mathcal{M}\setminus \{M\}} 
\renewcommand{\L}{\mathcal{L}} 
\renewcommand{\b}{b_{anc}} 
\renewcommand{\dj}{d_{anc,j}} 
\newcommand{\Le}{L_{anc}}  
\newcommand{\Ue}{U_{anc}} 
\newcommand{\deltat}{\Delta t} 

\renewcommand{\Re}{\mathbb{R}}

\begin{document}
\vspace*{0.2in}

\begin{flushleft}
{\Large
\textbf\newline{Optimized Treatment Schedules for Chronic Myeloid Leukemia} 
}
\newline
\\

Qie He\textsuperscript{1},
Junfeng Zhu\textsuperscript{1},
David Dingli\textsuperscript{2},
Jasmine Foo\textsuperscript{3**},
Kevin Leder\textsuperscript{2*},
\\
\bigskip
\textbf{1} Department of Industrial and Systems Engineering, University of Minnesota, Minneapolis, MN, USA
\\
\textbf{2} Department of Hematology, Mayo Clinic, Rochester, MN, USA
\\
\textbf{3} Department of Mathematics, University of Minnesota, Minneapolis, MN
\\
\bigskip

%
%





* lede0024@umn.edu@umn.edu, **jyfoo@math.umn.edu

\end{flushleft}
\section*{Abstract}
Over the past decade, several targeted therapies (e.g. imatinib, dasatinib, nilotinib) have been developed to treat Chronic Myeloid Leukemia (CML).  Despite an initial response to therapy,  drug resistance remains a problem for some CML patients.  Recent studies have shown that resistance mutations that preexist treatment can be detected in a substantial number of patients, and that this may be associated with eventual treatment failure.  One proposed method to extend treatment efficacy is to use a combination of multiple targeted therapies.  However, the design of such combination therapies (timing, sequence, etc.) remains an open challenge. In this work we mathematically model the dynamics of CML response to combination therapy and analyze the impact of combination treatment schedules on treatment efficacy in patients with preexisting resistance. We then propose an optimization problem to find the best schedule of multiple therapies based on the evolution of CML according to our ordinary differential equation model. This resulting optimization problem is nontrivial due to the presence of ordinary different equation constraints and integer variables. Our model also incorporates realistic drug toxicity constraints by tracking the dynamics of patient neutrophil counts in response to therapy.  Using realistic parameter estimates, we determine optimal combination strategies that maximize  time until treatment failure.

\section*{Author Summary}
Targeted therapy using imatinib, nilotinib or dasatinib has become standard treatment for chronicle myeloid leukemia. A minority of patients, however, fail to respond to treatment or relapse due to drug resistance. One primary driving factor of drug resistance are point mutations within the driving oncogene. Laboratory studies have shown that different leukemic mutants respond differently to different drugs, so a promising way to improve treatment efficacy is to combine multiple targeted therapies. We build a mathematical model to predict the dynamics of different leukemic mutants with imatinib, nilotinib and dasatinib, and employ optimization techniques to find the best treatment schedule of combining the three drugs sequentially. Our study shows that the optimally designed combination therapy is more effective at controlling the leukemic cell burden than any monotherapy under a wide range of scenarios. The structure of the optimal schedule depends heavily on the mutant types present, growth kinetics of leukemic cells and drug toxicity parameters. Our methodology is an important step towards the design of personalized optimal therapeutic schedules for chronicle myeloid leukemia.

\section*{Introduction}
Chronic Myeloid Leukemia (CML) is an acquired hematopoietic stem cell disorder leading to the over-proliferation of myeloid cells and an increase in cellular output from the bone marrow that is often associated with splenomegaly. The most common driving mutation in CML is a translocation between chromosomes 9 and 22 that produces a fusion gene known as BCR-ABL. The BCR-ABL protein promotes proliferation and inhibits cell apoptosis of myeloid progenitor cells and thereby drives expansion of this cell population. By targeting the BCR-ABL oncoprotein, imatinib (brand name Gleevec) is able to induce a complete cytogenetic remission in the majority of chronic phase CML patients.  A minority of patients, however, either fail to respond or eventually develop resistance to treatment with imatinib \cite{druker2006five}. It is thought that a primary driver of this resistance to imatinib is point mutations within the BCR-ABL gene.  A recent study utilizing sensitive detection methods demonstrated that a small subset of these mutations may exist before the initiation of therapy in a significant fraction of patients, and that this status is correlated with eventual treatment failure~\cite{iqbal2013sensitive}.  Second generation agents such as dasatinib and nilotinib have been developed and each has shown efficacy against various common mutant forms of BCR-ABL. This leads to the observation that the various mutant forms of BCR-ABL result in CML that have unique dynamics under therapy, and that combinations of these inhibitors may be necessary to effectively control a rapidly evolving CML population.  Patients with CML often die due to transformation of the disease into an acute form of leukemia known as blast crisis. It has been shown that blast crisis is due to the accumulation of additional mutations in CML progenitor cells~\cite{jamieson2004granulocyte}. 

The goal of this work is to leverage the differential responses of CML mutant strains to design novel sequential combination treatment schedules using dasatinib, imatinib and nilotinib that optimally control leukemic burden and delay treatment failure due to pre-existing resistance.  We develop and parametrize a mathematical model for the evolution of both wild-type (WT) CML and mutated (resistant) CML cells in the presence of each therapy. Then we formulate the problem as a discrete optimization problem in which a sequence of monthly treatment decisions is optimized to identify the temporal sequence of imatinib, dasatinib, and nilotinib administration that minimizes the total  CML cell population over a long time horizon. 


There has been a significant amount of work done in the past to mathematically model CML. For example, in  \cite{FoKeCl91} the authors developed a system of ordinary differential equations (ODEs) that model both the normal progression from stem cell to mature blood cells and abnormal progression of CML.  A hierarchical system of differential equations was used to model the response of CML cells to imatinib therapy in \cite{MiIwHu05}; this model fit the biphasic nature of decline in BCR-ABL positive cells during imatinib treatment. In \cite{leder2011fitness} the authors investigated the number of different resistant strains present in a newly diagnosed chronic phase CML patient. An optimal control approach was utilized to optimize imatinib scheduling in \cite{AiBe10}.  Particularly relevant to our work is \cite{komarova2009combination, KaKo10} where the authors investigated simultaneous continuous administration of dasatinib, nilotinib and imatinib; in particular, they explored the minimal number of drugs necessary to prevent drug resistance.  In the current work, we focus on understanding the optimal administration schedule of multiple therapies to prevent resistance, and studying the impact of toxicity constraints on optimal scheduling.  Since several of the available  tyrosine kinase inhibitors (TKI) share similar toxicities (in particular neutropenia, see e.g., \cite{quintas2004granulocyte,guilhot2007dasatinib,swords2009nilotinib}) combining them together can lead to elevated risk of adverse events.  Thus we consider sequential combination therapies in which only one TKI may be administered at a time. Moreover, it has been shown that the risk of treatment failure and blast crisis are highest within the first 2 years from diagnosis~\cite{druker2006five}. Therefore it is possible that optimized, sequential single agent therapy may be sufficient to minimize the risk of treatment failure.   Allowing only one treatment at a time leads to a complex, time-dependent discrete optimization problem. 

Another line of research closely related to the current work is the use of optimal control techniques in the design of optimal temporally continuous drug concentration profiles (see, e.g., review articles \cite{Swan90, Shi14} and the textbook \cite{MartinTeo93}). In this field the tools of optimal control such as the Pontryagin principle and the Euler-Lagrange equations are used to find drug concentration profiles that result in minimal tumor cell populations under toxicity constraints. Particularly relevant to the current work is \cite{WeZeNo97} where the authors searched for optimal anti-HIV treatment strategies.  They dealt with the similar problem of treating heterogeneous populations with multiple drugs. One major drawback of these works is the fact that it is nearly impossible to to achieve a specific optimal continuous drug-concentration profile in patients, since drug concentration over time is a combined result of a treatment schedule (e.g. sequence of discrete oral administrations) and pharmacokinetic processes in the body including metabolism, elimination, etc.  Thus the clinical utility of an optimal continuous drug concentration profile is limited.  In contrast to these previous works, here we model the optimization problem as a more clinically realistic sequence of monthly treatment decisions. Imposition of this fixed discrete set of decision times leads to a challenging optimization problem. Such dynamical systems are referred to as `switched nonlinear systems' in the control community~\cite{liberzon2012switching}, and our problem additionally imposes fixed switching times. In this work we will leverage the system structure and tools from mixed-integer linear optimization~\cite{nemhauser1999integer} to solve this problem numerically, resulting in optimal therapy schedules that are easy to implement in practice.

\section*{Computational framework}
\subsection*{Model of CML dynamics}
\label{sec:Model}

We consider an ODE model of the differentiation hierarchy of hematopoietic cells, adapted from~\cite{MiIwHu05, Foo_PlosCBCML, olshen2014dynamics}. Stem cells (SC) on top of the hierarchy give rise to progenitor cells (PC), which produce differentiated cells (DC), which in turn produce terminally differentiated cells (TC). This differentiation hierarchy applies to both normal and leukemic cells~\cite{strife1988biology}. We consider in our model leukemic WT cells as well as pre-existing BCR-ABL mutant cell types.  We use type 1, type 2, and type $i$ ($3\le i \le n$) cells to denote normal, leukemic WT, and $(n-2)$ leukemic mutant cells; layer 1, 2, 3, 4 cells to denote SC, PC, DC, and TC; and drug 0, 1, 2, 3 to denote a drug holiday, nilotinib, dasatinib, and imatinib, respectively. Let $x_{l,i}(t)$ denote the abundance of type $i$ cell at layer $l$ and time $t$, and $x(t)=(x_{l,i}(t))$ be the vector of all cell abundance at time $t$. If drug $j\in \{0,1,2,3\}$) is taken from month $m$ to month $m+1$, then the cell dynamics are modeled by the following set of ODEs.
\begin{subequations} \label{eq:odeconcise:j}
\begin{align}
\dot{x}(t)=f^j(x(t)), \ & t\in [m\deltat, (m+1)\deltat],\\
x(m\deltat) = x^m, \ &
\end{align}
\end{subequations}
for some function $f^j$, where $\deltat=30$ days and $x^m$ is the cell abundance at the beginning of month $m$. The concrete form of function $f^j$ under drug $j$ is described as follows. 
\begin{subequations} \label{eq:hierarchical}
\begin{align}
\text{SC level} \qquad & \dot{x}_{1,i}=(b^j_{1,i}\phi_i - d^j_{1,i})x_{1,i}, \; i=1,\ldots,n\\
\text{PC level} \qquad & \dot{x}_{2,i}= b^j_{2,i} x_{1,i} - d^j_{2,i}x_{2,i}, \; i=1,\ldots,n \\
\text{DC level} \qquad & \dot{x}_{3,i}= b^j_{3,i} x_{2,i} - d^j_{3,i}x_{3,i}, \; i=1,\ldots,n \\
\text{TC level} \qquad & \dot{x}_{4,i}= b^j_{4,i} x_{3,i} - d^j_{4,i}x_{4,i}, \; i=1,\ldots,n.
\end{align}
\end{subequations}
Here we describe the function of each parameter of this model.  For a detailed discussion of how these parameters were estimated from biological data, please see section  \ref{sec:PARAM} of the  Appendix.  Type $i$ stem cells divide at rate $b^j_{1,i}$ per day under drug $j$. The production rates of type $i$ progenitors, differentiated cells, and terminally differentiated cells under drug $j$ are $b^j_{l,i}$ per day for $l=2,3,4$, respectively. The type $i$ cell at layer $l$ dies at rate $d^j_{l,i}$ per day under drug $j$, for each $i$, $l$ and $j$. The competition among normal and leukemic stem cells is modeled by the density dependence functions $\phi_i(t)$, where $\phi_i(t) =1/(1+p_i\sum_{i=1}^nx_{1,i}(t))$ for each $i$; these functions ensure that the normal and leukemic stem cell abundances remain the same once the system reaches a steady state. The parameter $p_1$ (resp. $p_2$) is computed from the equilibrium abundance of normal (resp. leukemic WT) stem cells assuming only normal (resp. leukemic WT) cells are present, and we set $p_i=p_2$ for each $i\ge 3$. In particular, $p_1=(b^0_{1,1}/d^0_{1,1}-1)/K_1$ and $p_2 = (b^0_{1,2}/d^0_{1,2}-1)/K_2$, with $K_1$ (resp. $K_2$) being the equilibrium abundance of normal (resp. leukemic WT) stem cells assuming only normal (resp. leukemic WT) cells are present.

\subsection*{Toxicity modeling}
\label{subsec:ANC}

 One of the most common side effects of TKIs in CML is neutropenia, or the condition of abnormally low neutrophils in the blood.  Neutropenia is defined in terms of the absolute neutrophil count (ANC).  To incorporate toxicity constraints we develop a model of the dynamics of the patient's ANC in response to each therapy schedule.  We then constrain our optimization problem by considering only schedules during which the patient's ANC stays above an acceptable threshold level $\Le$. Typically, ANC at diagnosis is within normal limits (between $1500-8000/\mathrm{mm}^3$); thus we set each patient's initial ANC to be $3000/\mathrm{mm}^3$.  Treatment with imatinib, dasatinib and nilotinib all result in reduction of the ANC at varying rates.  Neutropenia is defined as an ANC level below $\Le = 1000/\mathrm{mm}^3$.  If a patient's ANC falls below the threshold, a drug holiday is required at the next monthly treatment decision stage.  During a drug holiday, ANC level will recover back to safe levels.
 
To model this process, we assume the patient's ANC decreases at rate $d_{anc,j}$ per month taking drug $j$, for $j=1,2,3$. During a drug holiday, ANC increases at rate $b_{anc}$ per month but never exceeds the normal level of $\Ue = 3000/\mathrm{mm}^3$. More specifically, let $y^m$ denote the ANC level at the beginning of month $m$ and the binary variable $z^{m,j}$ indicate whether drug $j$ is taken in month $m$ or not, i.e., $z^{m,j}=1$ (resp. $z^{m,j}=0$) indicates drug $j$ is (resp. not) taken in month $m$. The kinetics of ANC is modeled through the truncated linear function
\begin{equation*}
 y^{m+1}=r(y^m, z^{m,0},z^{m,1}, z^{m,2}, z^{m,3})=\min\{y^m + \b z^{m,0} - \sum_{j\in \{1,2,3\}} d_{anc,j}z^{m,j}, \Ue\}.
\end{equation*}
For example, after a patient with ANC level $y^m$ takes a drug holiday in month $m$, her ANC level at the beginning of month $m+1$ becomes $y^m+\b$ if $y^m+\b$ is not higher than the normal level $\Ue$, or $\Ue$ if $y^m+\b$ exceeds $\Ue$. If the patient instead takes nilotinib in month $m$, then her ANC level at the beginning of month $m+1$ becomes $y^m  - d_{anc,1}$.  The parameters governing ANC rate of change are provided in section \ref{sec:PARAM} of  the Appendix.

\section*{Treatment optimization problem} \label{sec:optmodel}
Assume the initial population of each cell type is known. Our goal is to select a treatment plan to minimize the tumor size at the end of the planning horizon subject to certain toxicity constraints. We call this the Optimal Treatment Plan problem (OTP). Each treatment plan is completely characterized by a temporal sequence of monthly treatment decisions over a long time horizon.  Between each monthly treatment decision, the dosing regimen is identical from day to day.  The standard regimens for each drug, which we will utilize throughout the work, are $300$mg twice daily for nilotinib, $100$mg once daily for dasatinib, and $400$mg once daily for imatinib~\cite{o2012chronic}.   For example, let $1$ denote nilotinib, 2 - dasatinib, 3 - imatinib, and 0 - drug holiday.  Then the sequence $\{1,1,\ldots,1\}$ represents that the patient takes the standard nilotinib regimen, $300$mg twice daily, every day, every month.  The sequence $\{2, 0, 2, 0, \ldots \}$ represents that the patient alternates between the standard dasatinib regimen, $100$mg once daily,  and a drug holiday on alternate months.

We introduce the binary decision variables $z^{m,j}$ to indicate whether drug $j$ is taken in month $m$ or not, for each $j=0,1,2,3$ and $m=0,1,\ldots,M-1$. An assignment of values to all $z^{m,j}$ variables that satisfy all constraints in the optimization model gives a feasible treatment plan.

\subsection*{The optimization problem}
Note the total leukemic cell abundance at day $t$ is given by $\sum_{l \ge 1}\sum_{i\ge 2} x_{l,i}(t)$.  The OTP can be formulated as the following mixed-integer optimization problem with ODE constraints.
\begin{subequations} \label{eq:OTP}
\begin{align}
	\min \ & \sum_{l \ge 1}\sum_{i\ge 2} x_{l,2}(M\deltat)  \label{eq:OTP:obj}\\
	\text{s.t.} \ & \dot{x}(t) = \sum_{j=0}^3 z_{m,j}f^j(x(t)) , & t\in [m\deltat, (m+1)\deltat], m=0,1,\ldots,M-1, \label{eq:OTP:ode}\\
		& \sum_{j=0}^3 z^{m,j} = 1,  &m=0,1,\ldots,M-1, \label{eq:OTP:card}\\
		& y^{m+1} = r(y^m, z^{m,0}, z^{m,1}, z^{m,2}, z^{m,3}), &m=0,1,\ldots,M-1, \label{eq:OTP:tox1}\\		
		& y^m \ge \Le, &m=0,1,\ldots,M, \label{eq:OTP:tox2}\\
		& z^{m,j} \in \{0,1\}, & j=0,1,2,3, m=0,1,\ldots,M-1 \label{eq:OTP:binary}\\ 
		& x(0) = x^0, \;  y^0 \text{ is given}.
\end{align}
\end{subequations}
To summarize the previous display, in equation~\eqref{eq:OTP:obj} we state that our objective is to minimize the leukemic cell population at the end of the treatment horizon. In equation~\eqref{eq:OTP:ode} we stipulate that the cell dynamics are governed by the system of differential equations given by \eqref{eq:hierarchical}. Together~\eqref{eq:OTP:card} and~\eqref{eq:OTP:binary} stipulate that during each time period we administer either one drug or no drug. Equations ~\eqref{eq:OTP:tox1} and~\eqref{eq:OTP:tox2} reflect the toxicity constraints described above.

The OTP problem is a mixed-integer nonlinear optimization problem, in which some constraints are specified by the solution to a nonlinear system of ODEs \eqref{eq:OTP:ode}. This optimization problem is beyond the ability of state-of-the-art optimization software. However, if we assume the TKI therapies do not affect the stem cell compartment, then it is possible to handle the ODE constraints numerically. This is because the non-linearities in the ODE model are only present in the stem cell compartment, and the remaining compartments are modeled by linear differential equations. Thus we are able to build a refined linear approximation to the ODE constraints (see Section \ref{sec:LIN_ODE} of the Appendix), and recast the problem as a mixed-integer linear optimization problem (see Section \ref{sec:MILO} of the Appendix). 

\subsection*{A quick reference for notation}
Below we summarize our notation for the ease of the reader.
\begin{itemize}
	\item $\I=\{1,2,\ldots, n\}$: the set of cell types. Type 1 denotes normal cells, type 2 denotes leukemic WT cells, and type $i$ ($3\le i\le n$) denotes one type of leukemic mutants.	
	\item $\L=\{1,2,\ldots,L\}$: the set of cell layers. We have $L=4$, and layer 1, 2, 3 and 4 denotes SC, PC, DC, and TC, respectively.
	\item $\J=\{0,1,2,\ldots, J\}$: The set of drugs for CML. We have $J=3$, drug 0 refers to a drug holiday, and drug 1 to drug 3 refers to nilotinib, dasatinib, and imatinib, respectively.
	\item $\M=\{0,1,\ldots,M\}$: the set of months for treatment. 
	\item $\deltat$: the duration during which a patient takes one drug before deciding to switch to another drug or take a drug holiday. We set $\deltat = 30$ days.
	\item $K_1$: the equilibrium abundance of normal stem cells when only normal cells are present.
	\item $K_2$: the equilibrium abundance of leukemic WT stem cells when only leukemic cells are present.
	\item $b^j_{l,i}$: the production rate of type $i$ cell at layer $l$ under drug $j$.	
	\item $d^j_{l,i}$: the death rate of type $i$ cell at layer $l$ under drug $j$. 
	\item $\b$: the average increase ($/\mathrm{mm}^3$) of the ANC in a patient without any drug after time $\deltat$.
	\item $\dj$: the average decrease ($/\mathrm{mm}^3$) of ANC in a patient under drug $j$ after time $\deltat$.	
	\item $\Le$: The lower limit of the ANC. We assume that the patient develops neutropenia if the ANC is less than $\Le$, at which a drug holiday needs to be taken.
	\item $\Ue$: The normal level of ANC.
\end{itemize}

\section*{Results}

In this work we consider the dynamics of CML response to single-agent and combination schedules utilizing the standard therapies imatinib, dasatinib and nilotinib.

\subsection*{Evolution of preexisting BCR-ABL mutants under standard monotherapy}
\label{subsec:stand_sim}
We first utilize the model to demonstrate the dynamics of CML populations with preexisting BCR-ABL mutations under monotherapy with the standard therapies imatinib, dasatinib and nilotinib.  Recall that the standard dosing regimens are $300$mg twice daily for nilotinib, $100$mg once daily for dasatinib, and $400$mg once daily for imatinib~\cite{o2012chronic}.  Growth rate parameters for each cell type in the model are estimated using \emph{in vitro} IC50 values reported in~\cite{redaelli2009activity} for each drug.  The initial cell populations at the start of therapy are derived by running the model starting from clonal expansion of a single leukemic cell in a healthy hematopoietic system at equilibrium \cite{foo2009eradication} until CML detection (when the total leukemic burden reaches approximately $10^{12}$ cells~\cite{holyoake2002elucidating}).  At this point the total cell burden is 2-3 times the normal cell burden in a healthy individual and thus the total leukemic cells make up approximately $77\%$ of the total cell population; this is consistent with clinical reports \cite{dingli2008chronic}. Details on deriving the initial cell abundances at diagnosis are provided in  section \ref{sec:PARAM} of  the Appendix.

In the first example we consider a patient harboring a low level of the BCR-ABL mutant F317L (which is resistant to dasatinib) before the initiation of TKI therapy. The initial population conditions are given in Table~\ref{tab:p1:F317L} with the leukemic WT and F317L cells taking up $95\%$ and $5\%$ of the leukemic cells, respectively.
\begin{table}[htb]
\centering
\caption{{\bf The initial cell abundance.}} 
\begin{tabular}{cccc}
\hline
 & normal cell & Wild-type & F317L \\
\hline
SC & $7.34 \times 10^{4}$ & $2.80 \times 10^{5}$ & $1.48 \times 10^{4}$ \\
PC & $1.61 \times 10^{7}$ & $3.87 \times 10^{7}$ & $2.04 \times 10^{6}$ \\			
DC & $3.24 \times 10^{9}$ & $1.03 \times 10^{10}$ & $5.40 \times 10^{8}$\\
TC & $3.24 \times 10^{11}$ & $1.03 \times 10^{12}$ & $5.40 \times 10^{10}$ \\
\hline
\end{tabular}
\label{tab:p1:F317L}
\end{table}

\begin{figure}[htb] 
 \centering
 \includegraphics[scale=0.9]{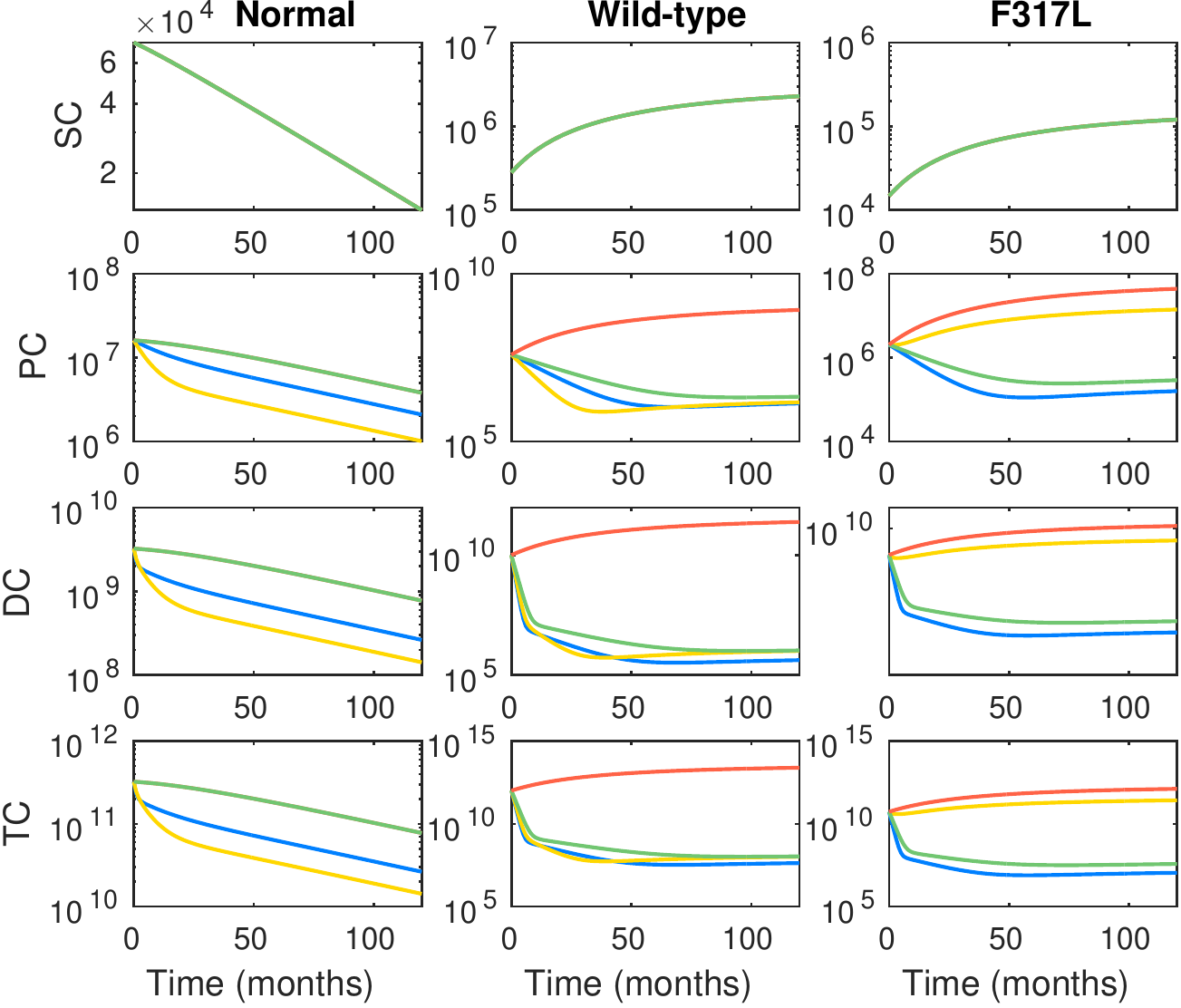}
\caption{{\bf Long term dynamics of healthy, WT leukemic and F317L mutant leukemic cell populations under treatment with standard regimen monotherapy nilotinib (blue), dasatinib (yellow), imatinib (green) and no drug (orange).} The dynamics of healthy normal cells with mono imatinib (green) and no drug (orange) coincide. Initial conditions are provided in Table~\ref{tab:p1:F317L} and parameter choices are provided in Appendix \ref{sec:PARAM}. }
\label{fig:nilotinib:p2}
\end{figure}

We plot in Fig~\ref{fig:nilotinib:p2} the cell dynamics over 120 months for four treatment plans: (1) nilotinib monotherapy (2) dasatinib monotherapy, (3) imatinib monotherapy, (4) no therapy - control.  We observe that as predicted, the disease burden responds well to  imatinib and nilotinib; the percentage of cancerous cells after a 24 month treatment drops to $0.19\%$ with nilotinib and $0.26\%$ with imatinib, respectively. However, the F317L mutant population is fairly resistant to dasatinib; we observe that the percentage of cancerous cells after 24 months is $58.1\%$ with dasatinib and $95.4\%$ with no treatment.  Over the 120 month period dasatinib treatment provides only modest improvement over the `no drug' option in controlling the F317L population; however, dasatinib remains quite effective in controlling the WT leukemic population.  It is interesting to note that overall, nilotinib is the most effective in controlling both the WT and F317L leukemic populations.  However, nilotinib also negatively impacts the healthy cell population more severely than imatinib, which is slightly less effective in controlling the leukemic populations.  This suggests that some trade-offs between these drugs exist, and these trade-offs may be exploited in designing combination therapies.

In the next example we consider a patient with BCR-ABL mutant type M351T preexisting therapy.    In contrast to the previous example, this commonly occurring mutant has been found to be partially sensitive in varying degrees to all three therapies.  The initial conditions are given in Table~\ref{tab:p1:M351T}.   Once again we have assumed that WT and M351T cells take up $95\%$ and $5\%$ of total leukemic cells, respectively.
\begin{table}[htb]
\centering
\caption{{\bf The initial cell abundance.}} \label{tab:p1:M351T}
\begin{tabular}{cccc}
\hline
 & Normal cell & Wild-type & M351T \\
\hline
SC & $7.34 \times 10^{4}$ & $2.80 \times 10^{5}$ & $1.48 \times 10^{4}$ \\
PC & $1.61 \times 10^{7}$ & $3.87 \times 10^{7}$ & $2.04 \times 10^{6}$ \\			
DC & $3.24 \times 10^{9}$ & $1.03 \times 10^{10}$ & $5.40 \times 10^{8}$\\
TC & $3.24 \times 10^{11}$ & $1.03 \times 10^{12}$ & $5.40 \times 10^{10}$ \\
\hline
\end{tabular}
\end{table}

\begin{figure}[htb] 
 \centering
 \includegraphics[scale=0.9]{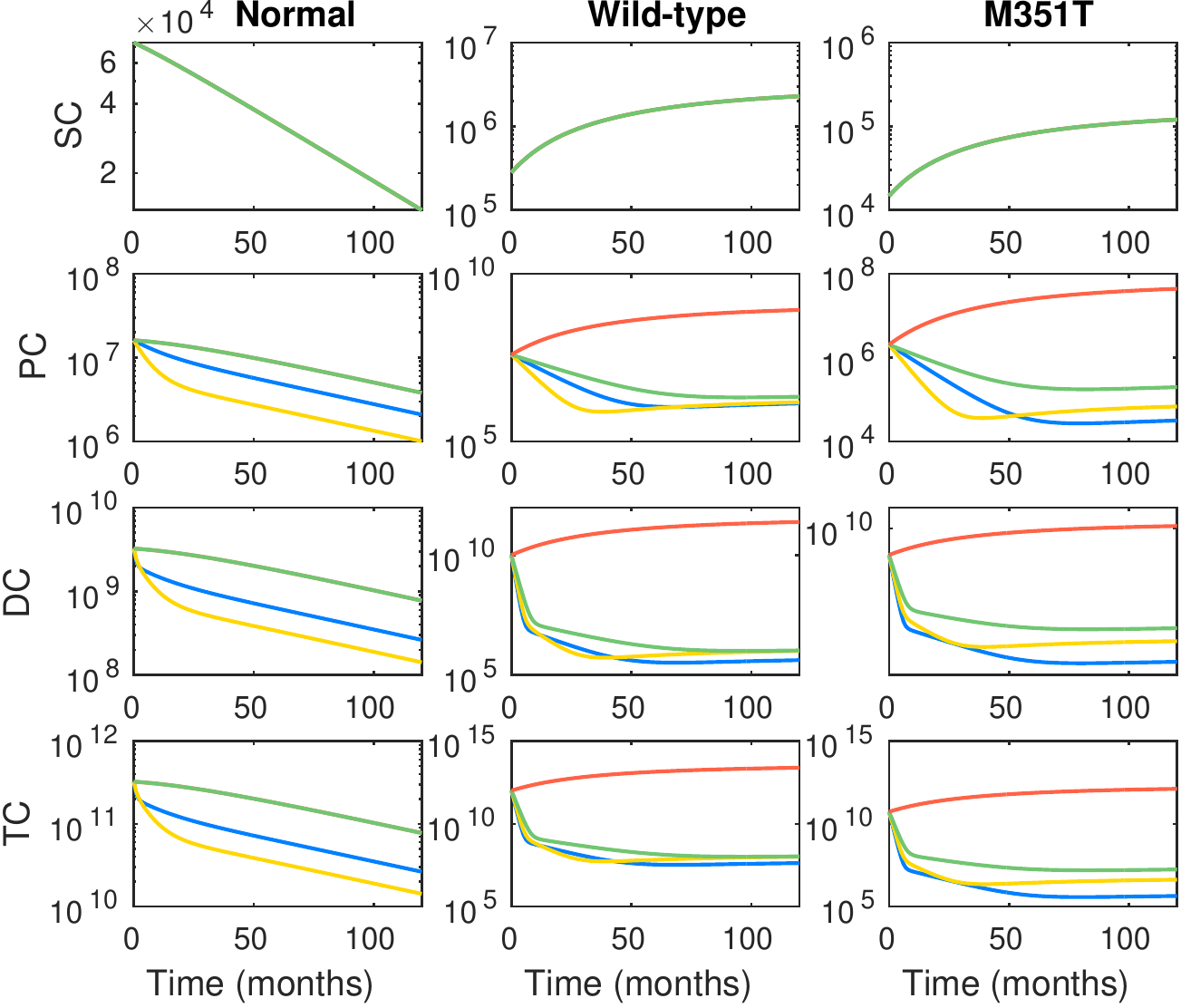}
  \caption{{\bf Long term dynamics of healthy, WT leukemic and M351T mutant leukemic cell populations under treatment with standard regimen monotherapy nilotinib (blue), dasatinib (yellow), imatinib (green) and no drug (orange).} The dynamics of healthy normal cells with mono imatinib (green) and no drug (orange) coincide. Initial conditions are provided in \ref{tab:p1:F317L} and parameter choices are provided in Appendix \ref{sec:PARAM}. }
\label{fig:nilotinib}
\end{figure}

In Fig~\ref{fig:nilotinib} the cell dynamics over 120 months for the four standard treatment plans are plotted:  (1) nilotinib monotherapy (2) dasatinib monotherapy, (3) imatinib monotherapy, (4) no therapy - control. Since the M351T mutant is responsive to each drug in contrast to the previous example, the percentage of cancerous cells after a 24 month treatment drops to $0.18\%$ with nilotinib, $0.18\%$ with dasatinib, and $0.25\%$ with imatinib, respectively. Without treatment, the percentage of cancerous cells after 24 months is $95.4\%$.  Here, we observe that although nilotinib is more effective than dasatinib in controlling the total mutant M351T burden, the effect is reversed in the progenitor population.  Higher levels of stem and progenitor populations will lead to faster rebound during treatment breaks, suggesting another trade-off to consider in the combination setting.  

\subsection*{Optimization of combination therapies}

We next solve the discrete optimization problem to identify sequential combination therapies utilizing imatinib, dasatinib and nilotinib to optimally treat CML patients with preexisting BCR-ABL mutations.  We consider schedules in which a monthly treatment decision is made between one of four choices: imatinib, dasatinib, nilotinib, and drug holiday.  During months in which one of the three drugs is administered, the dosing regimen is fixed at $300$mg twice daily for nilotinib, $100$mg once daily for dasatinib, and $400$mg once daily for imatinib.  In the following we optimize over feasible treatment decision sequences that result in a minimal leukemic cell burden after 3 years.  Each treatment plan is completely characterized by a temporal sequence of drugs over a long time horizon.  

\subsubsection*{Optimal therapy for preexisting M351T mutation, no toxicity constraints}
\label{sec:M3NoTox}

In our first example we assume that the mutant M351T preexists therapy.  For demonstration purposes no toxicity constraint is considered in this example. The initial cell populations are given in Table~\ref{tab:p1:M351T}. The remaining parameters are described in Section~\ref{sec:PARAM}. Note that WT and M351T leukemic cells comprise $95\%$ and $5\%$ of leukemic cells, respectively. The optimal schedule we obtain for this scenario is provided in Table \ref{tab:JF_b}. The proposed combination therapy is similar to the monotherapy using dasatinib, but switches to nilotinib towards the end of the 36 month time horizon.  We note that the optimization result is robust to changes in the initial abundance of the leukemic mutant cells; 
increasing the frequency of initial M351T mutants to $50\%$ of the leukemic population results in an almost identical optimal schedule (data not shown).  

\begin{table}[htb]
\begin{adjustwidth}{-0.5in}{0in}
\centering
\caption{{\bf Optimized treatment schedule for preexisting M351T.}}   
\begin{tabular}{cc}
\hline														
Optimal combination  & 2 2 2 2 2 2 2 2 2 2 2 2 2 2 2 2 2 2 2 2 2 2 2 2 2 2 2 2 2 2 2 1 1 1 1 1 \\
\hline
\end{tabular}
\begin{flushleft} Initial conditions are provided in Table~\ref{tab:p1:M351T}.  No toxicity constraints. Recall that 0 - Drug Holiday, 1 - Nilotinib, 2 - Dasatinib, and 3 - Imatinib.
\end{flushleft}
\label{tab:JF_b}
\end{adjustwidth}
\end{table}

\begin{figure}[htb]
\centering
 \includegraphics[scale=0.9]{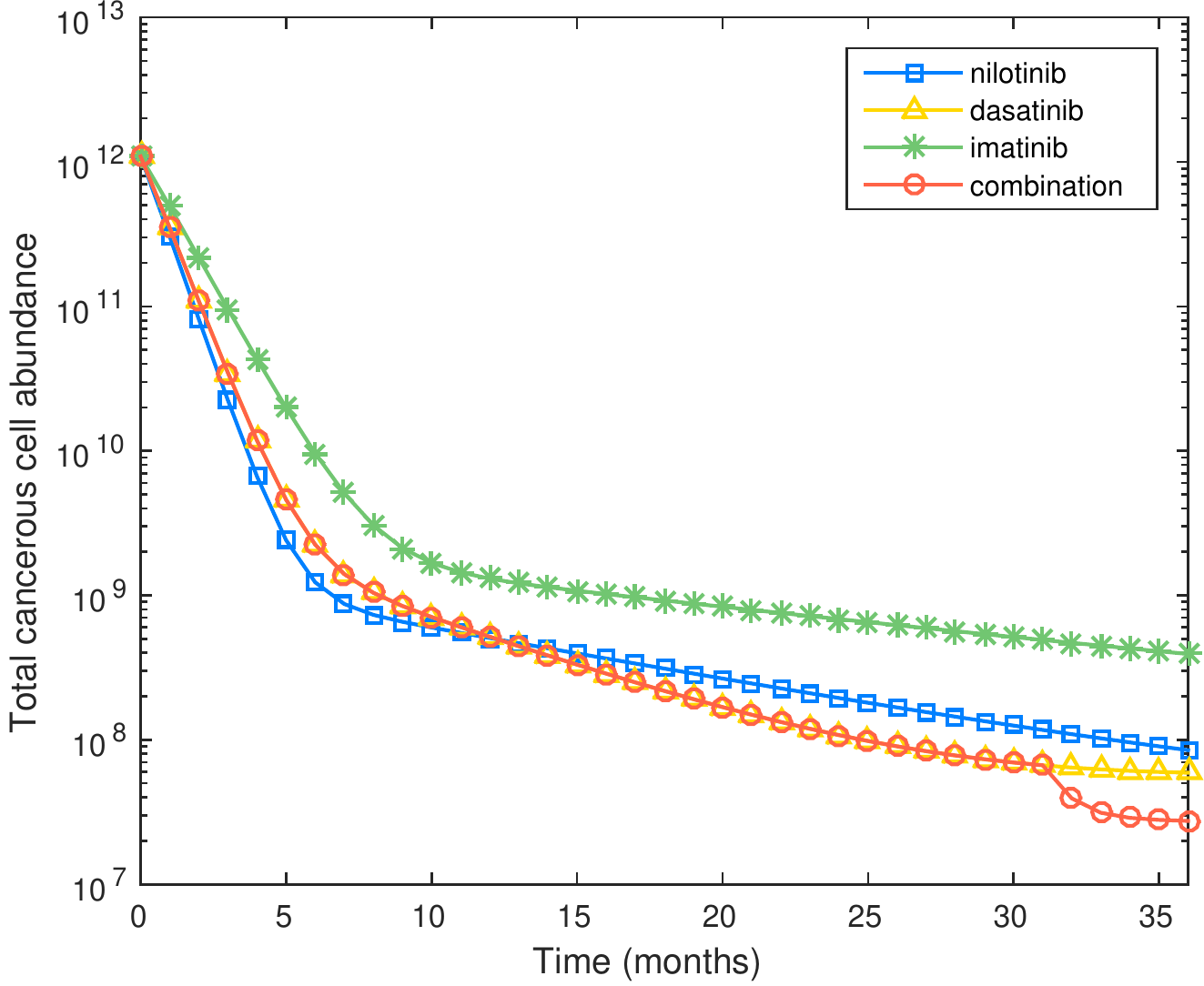}
\caption{{\bf Plot of cell number versus time for three monotherapies and optimal combination therapy for preexisting M351T, with no toxicity constraints.}}
\label{fig:opt}
\end{figure}

In Fig~\ref{fig:opt} we compare the performance of 4 different schedules including the optimized schedule. Amongst the four schedules tested the optimal schedule provides the lowest leukemic cell burden at the 36 month mark.
It is interesting to see that there is no single best drug. For monotherapy, it is best to use nilotinib if the treatment horizon is shorter than 12 months, and dasatinib if the treatment horizon longer than 12 months. The proposed combination therapy performs better than three monotherapies at the 36 month mark: the leukemic cell population at the end of 36 months is $2.75\times 10^7$ with the proposed combination therapy and $5.92 \times 10^7$ with dasatinib (the best monotherapy). We can see that the proposed optimal treatment schedule leads to more than $50\%$ reduction on final leukemic cell abundances over the best monotherapy. Figure~\ref{fig:opt} also shows that imatinib has less efficacy than nilotinib or dasatinib in reducing the leukemic cell burden when WT and M351T are present. An important question is, why does the optimal schedule take that specific form. In our parameter estimates (see Appendix \ref{sec:PARAM}) we see that dasatinib is better at killing progenitors than nilotinib, while nilotinib is better at killing differentiated cells. Thus the optimal schedule uses dasatinib at first to bring down the progenitor cell population, and then switches to nilotinib near the end of the treatment horizon to decrease the population of differentiated cells.  

{\it Optimal schedules robust to varying objective function and treatment length.}
We also consider an alternative objective function in which the goal is to minimize the average leukemic cell burden over the whole treatment horizon. Consider the scenario with M351T  mutation preexisting at initiation of therapy with no toxicity constraints again.  The altered objective function results in an optimal strategy of nilotinib monotherapy. To understand this, we note that for minimizing area under the cell population curve it is important to decrease the initial tumor population as quickly as possible. This tends to favor taking nilotinib the entire time since it leads to the quickest reduction in total tumor cell population, by reducing differentiated and therefore terminally differentiated cells.  We also ran optimization experiments to evaluate the impact of varying the length of treatment between 35 and 38 months; these resulted in very similar optimal schedules.

\subsubsection*{Optimal therapy for preexisting F317L mutation, no toxicity constraints}
Next we consider a patient with preexisting mutant F317L instead of M351T. According to the \emph{in vitro} IC50 value reported in~\cite{redaelli2009activity}, F317 is resistant to dasatinib, and moderately resistant to nilotinib and imatinib. The initial cell abundances are given in Table~\ref{tab:p1:F317L}; the mutant leukemic cells make up of $5\%$ of the total leukemic cells as in the baseline model except we replace mutant M351T with F317L. The proposed combination therapy is listed in Table~\ref{tab:p1:F317L:opt}, and for comparison the optimal therapy for the previous example where M351T preexisted therapy is also provided. Note that dasatinib is used in the first 9 months and nilotinib is used in the next 27 months in the presence of F317L. 
\begin{table}[htb]
\begin{adjustwidth}{-0.5in}{0in}
\centering
\caption{{\bf Optimal combination schedules.}} 
\begin{tabular}{cc}
\hline
M351T preexisting  & 2 2 2 2 2 2 2 2 2 2 2 2 2 2 2 2 2 2 2 2 2 2 2 2 2 2 2 2 2 2 2 2 1 1 1 1 \\
F317L preexisting  & 				 			2 2 2 2 2 2 2 2 2 1 1 1 1 1 1 1 1 1 1 1 1 1 1 1 1 1 1 1 1 1 1 1 1 1 1 1\\
\hline
\end{tabular}
\label{tab:p1:F317L:opt}
\end{adjustwidth}
\end{table}

\begin{figure}[htb]
\centering
 \includegraphics[scale=0.9]{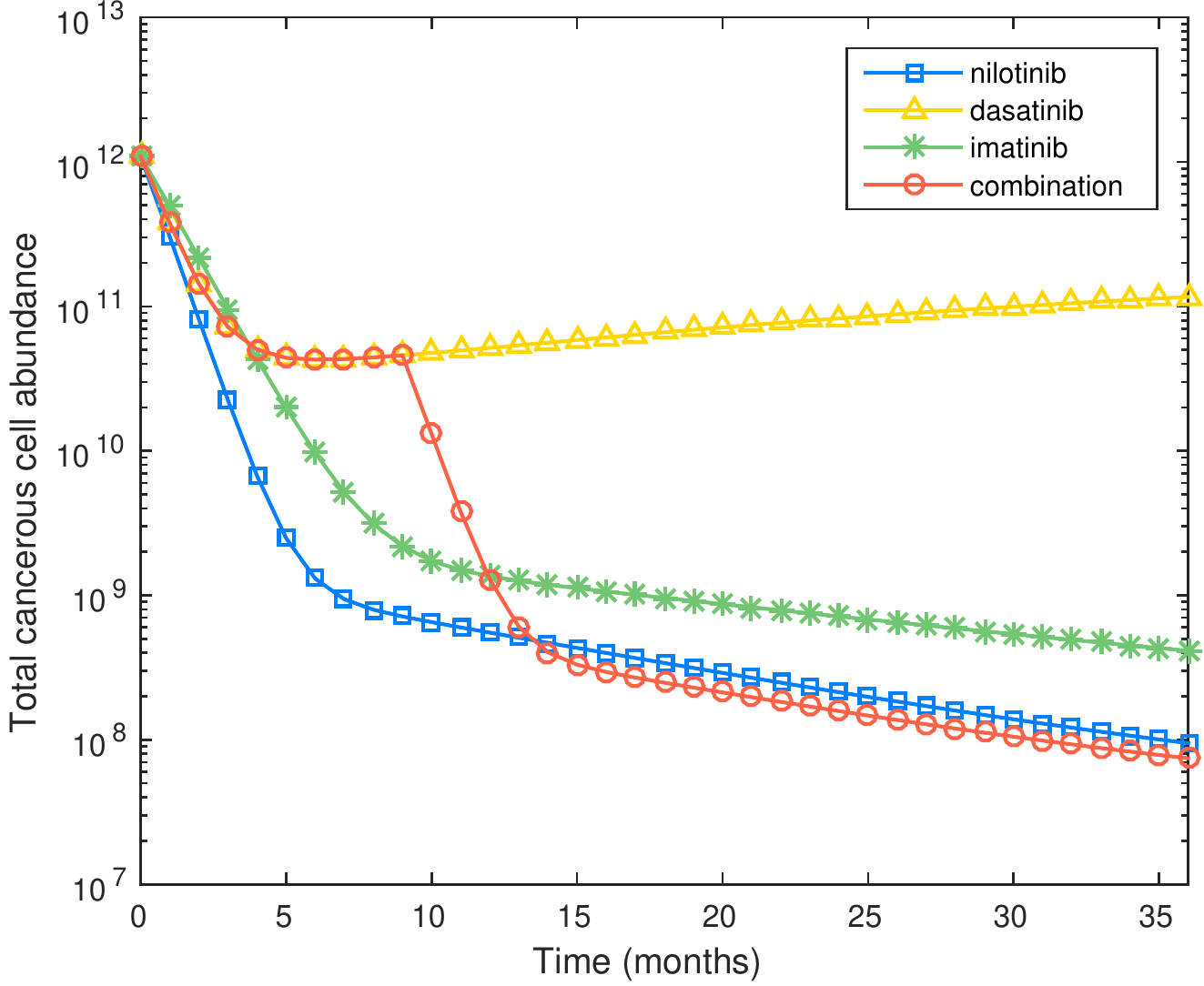}
 \caption{{\bf Plot of cell number versus time for three monotherapies and optimal combination therapy for preexisting F317L, with no toxicity constraints.}}
\label{fig:p3}
\end{figure}

In Fig~\ref{fig:p3} we show the comparison between proposed schedule and three different monotherapies. The final leukemic cell abundances are $7.46\times 10^7$ and $9.48\times 10^7$ under the propose schedule and monotherapy with nilotinib, respectively. The combination therapy performs better in reducing final leukemic cell population than three monotherapies, but the improvement is marginal in this case. Again the optimal schedule uses dasatinib to reduce the wild-type progenitor cell population, but switches to nilotinib much earlier to reduce the wild-type differentiated cell and F317L cell popoluations.

%
%

\subsubsection*{Incorporating toxicity constraints}
We next  study  how drug toxicity affects the optimal therapy, in particular with the drug toxicity constraint introduced in the \textbf{Toxicity Modeling}  subsection. Recall that the toxicity constraint prevents ANC from dipping below a threshold value $\Le$. The ANC decreases at a constant rate each month under each drug, and increases at a constant rate without drug. The ANC never exceeds the normal level of $\Ue$. We first assume that nilotinib has a higher toxicity than dasatinib, and dasatinib has a higher toxicity than imatinib. In particular, the monthly decrease rates of ANC for nilotinib, dasatinib, and imatinib are $350/\mathrm{mm}^3$, $300/\mathrm{mm}^3$, and $250/\mathrm{mm}^3$, respectively, and ANC increases by $2,000/\mathrm{mm}^3$ with one month drug holiday. 

We incorporated the toxicity constraints into the preexisting M351T mutant scenario described previously, i.e. initial cell populations are given in Table \ref{tab:p1:M351T}.   The three monotherapies and resulting optimal combination therapy are shown in Table~\ref{tab:tox1} below. Note that the proposed combination therapy is very close to the one described without toxicity constraints (i.e., Table \ref{tab:JF_b}), except now drug holidays are inserted to maintain the ANC level above $\Le$.
\begin{table}[htb]
\begin{adjustwidth}{-0.5in}{0in}
\centering
\caption{{\bf Treatment schedules with drug toxicities.}} 
\begin{tabular}{cc}
\hline
Nilotinib 	& 			1 1 1 1 1 0 1 1 1 1 1 1 0 1 1 1 1 1 1 0 1 1 1 1 1 0 1 1 1 1 1 1 0 1 1 1\\
Dasatinib 	&				2 2 2 2 2 2 0 2 2 2 2 2 2 2 0 2 2 2 2 2 2 0 2 2 2 2 2 2 2 0 2 2 2 2 2 2\\	
Imatinib 		&				3 3 3 3 3 3 3 0 3 3 3 3 3 3 3 3 0 3 3 3 3 3 3 3 3 0 3 3 3 3 3 3 3 3 0 3 \\
Combination &	2 0 2 2 2 2 2 2 0 2 2 2 2 2 2 0 2 2 2 2 2 2 0 2 2 2 2 2 2 0 2 2 1 1 1 1 \\
\hline
\end{tabular}
\label{tab:tox1}
\end{adjustwidth}
\end{table}

\begin{figure}[htb]
\centering
 \includegraphics[scale=0.9]{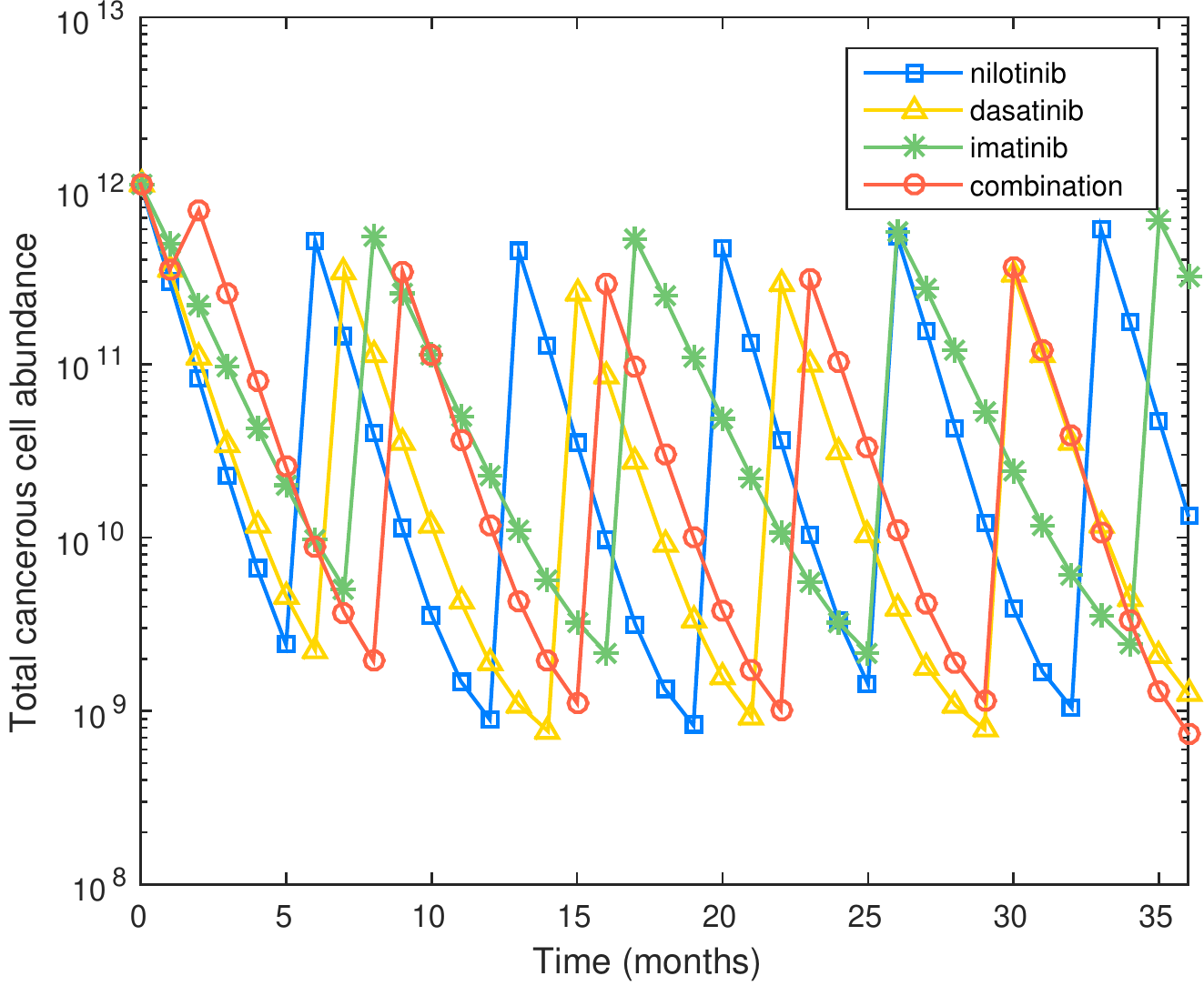}
 \caption{{\bf Plot of cell number versus time for three monotherapies and optimal combination therapy for preexisting M351T, incorporating toxicity constraints.}}
\label{fig:tox1}
\end{figure}

The cell dynamics of three monotherapies and the proposed combination therapy are given in Fig~\ref{fig:tox1}. It can be seen that after drug holidays, the total leukemic cell population almost returns to the level at the beginning of treatment. This indicates that a one month drug holiday may be too long for the patient.

\begin{figure}[htb]
\centering
 \includegraphics[scale=0.9]{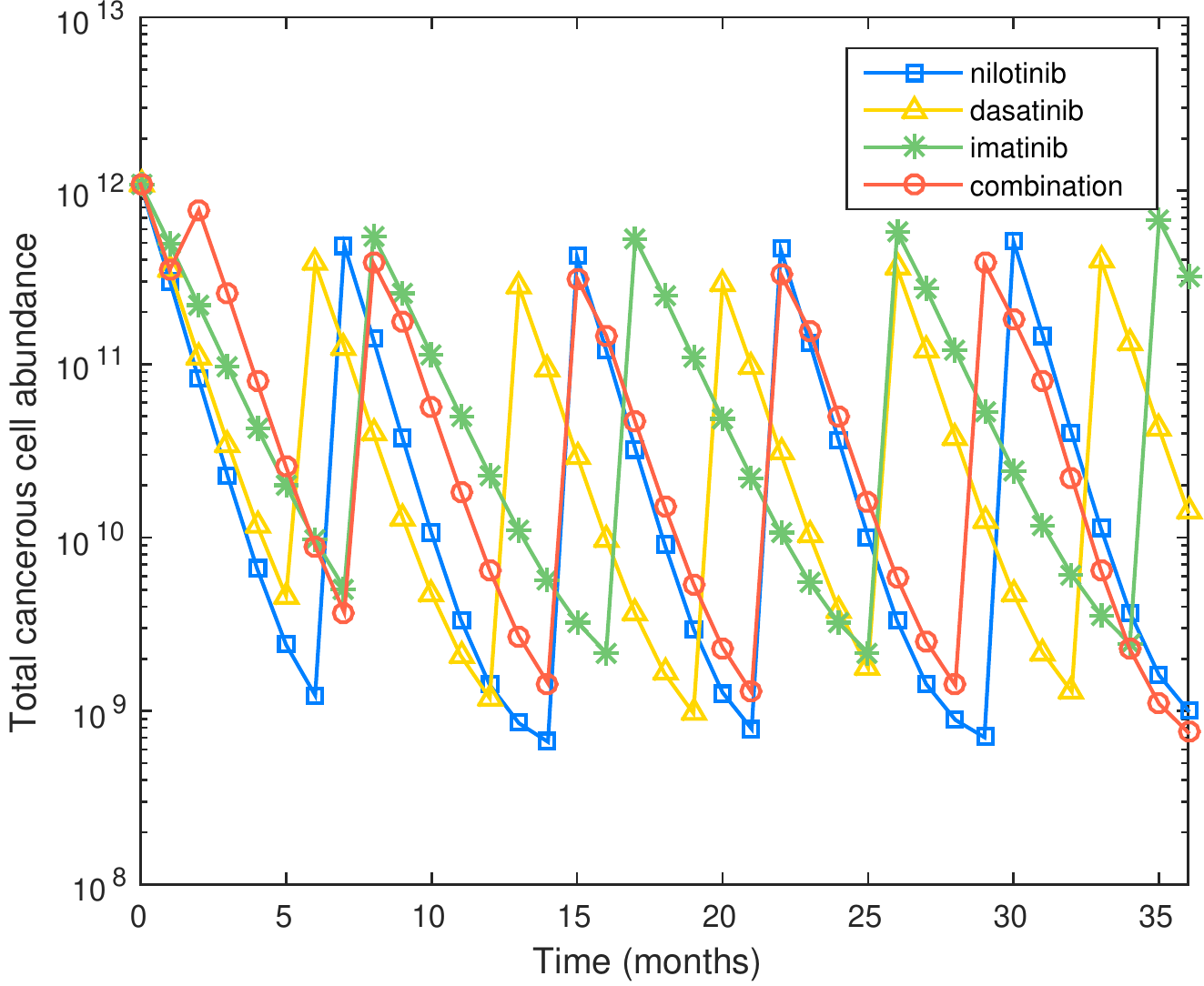}
 \caption{{\bf Plot of cell number versus time for three monotherapies and optimal combination therapy for preexisting M351T, incorporating toxicity constraints.} Here it is assumed that the ANC reduction rate during dasatinib treatment is higher than during nilotinib treatment.}
\label{fig:tox2}
\end{figure}

Since it is not clear whether nilotinib or dasatinib result in higher toxicity effects, we also switched the monthly ANC depletion rates to nilotinib - 300, dasatinib - 350, and imatinib - 250, so that dasatinib has the highest toxicity. Other conditions are kept the same. The recommended combination therapy is shown in Table~\ref{tab:tox2} below. Note that now imatinib is used more frequently, due to the increase in toxicity of dasatinib. We also compare the performance of the four different schedules in Fig~\ref{fig:tox2}.
\begin{table}[htb]
\begin{adjustwidth}{-0.5in}{0in}
\centering
\caption{{\bf The optimal treatment schedules with different drug toxicities.}} 
\begin{tabular}{cc}
\hline
Nilotinib$>$Dasatinib  & 						2 0 2 2 2 2 2 2 0 2 2 2 2 2 2 0 2 2 2 2 2 2 0 2 2 2 2 2 2 0 2 2 1 1 1 1 \\
Dasatinib$>$Nilotinib &  						2 0 2 2 2 2 2 0 3 2 2 2 2 2 0 3 2 2 2 2 2 0 3 2 2 2 2 2 0 3 3 1 1 1 1 1\\
\hline
\end{tabular}
\label{tab:tox2}
\end{adjustwidth}
\end{table}

\subsubsection*{Multiple mutants preexisting before the initiation of therapy}
Lastly we investigate how much gain can be expected from combination therapy if more than one mutant type preexists before initiation of therapy. We again consider an optimization model over a 36 month horizon. We assume mutants M351T and F317L preexist therapy at a low level (each consists of $5\%$ of the total leukemic cell population); the initial conditions are given in Table~\ref{tab:2mutants:init}. 
\begin{table}[htb]
\begin{adjustwidth}{-0.5in}{0in}
\centering
\caption{{\bf The initial cell abundance.}} 
\begin{tabular}{ccccc}
\hline
 & Normal cell & Wild-type & M351T & F317L\\
\hline
SC & $7.34 \times 10^{4}$ & $2.66 \times 10^{5}$ & $1.48 \times 10^{4}$ & $1.48 \times 10^{4}$\\
PC & $1.61 \times 10^{7}$ & $3.66 \times 10^{7}$ & $2.04 \times 10^{6}$ & $2.04 \times 10^{6}$\\			
DC & $3.24 \times 10^{9}$ & $9.72 \times 10^{9}$ & $5.40 \times 10^{8}$ & $5.40 \times 10^{8}$\\
TC & $3.24 \times 10^{11}$ & $9.72 \times 10^{11}$ & $5.40 \times 10^{10}$ & $5.40 \times 10^{10}$\\
\hline
\end{tabular}
\label{tab:2mutants:init}
\end{adjustwidth}
\end{table}
The recommended combination therapy is the same as the recommended therapy when only one mutant F317L is present. The result is reasonable since the F317L has higher resistance to our therapies, and thus has a more significant impact on the structure of the optimal treatment schedule.

\begin{figure}[htb]
\centering
\includegraphics[scale=0.9]{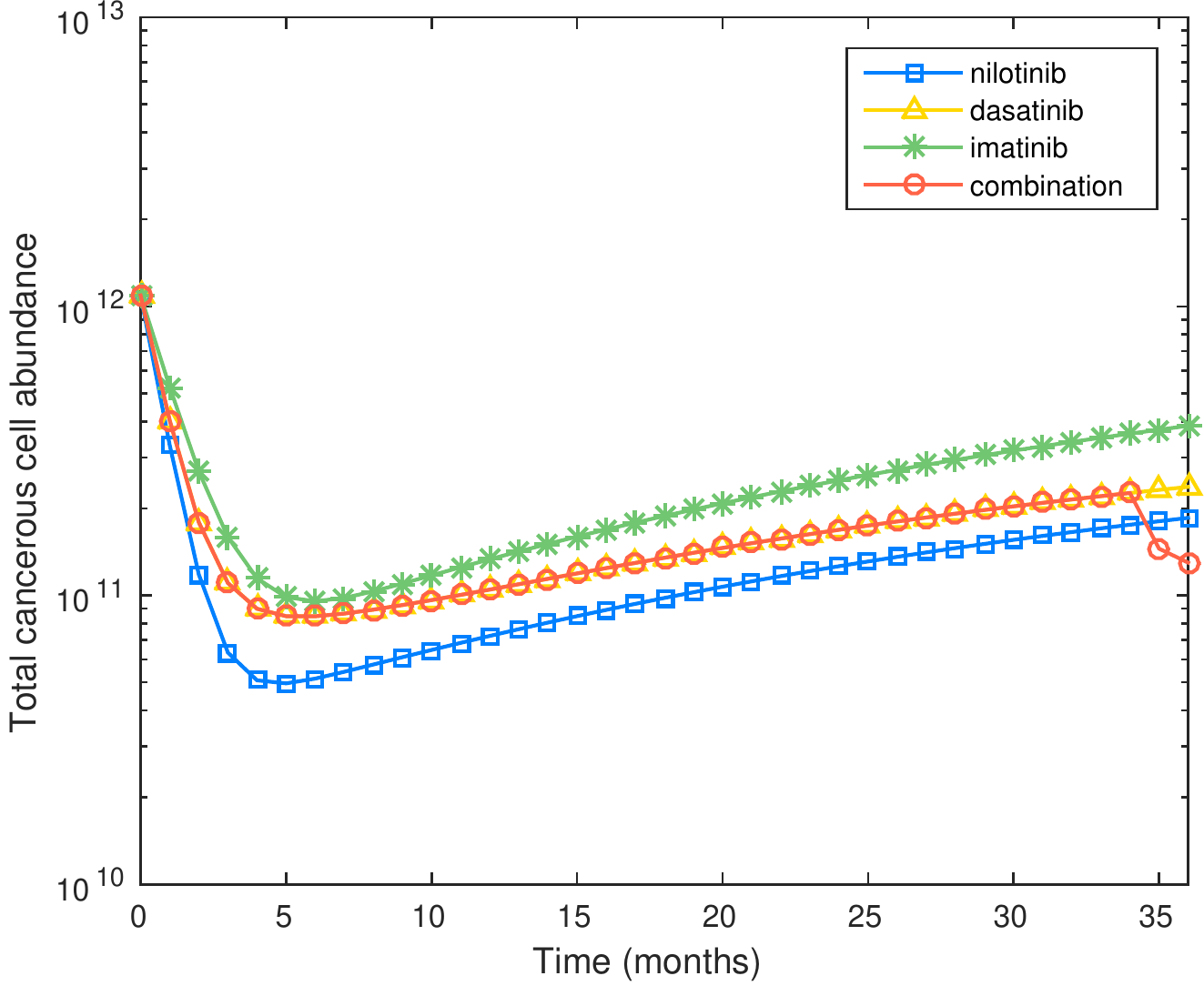}
 \caption{{\bf The cell dynamics with F317L and E255K preexisting therapy.}}
\label{fig:2mutants:therapy}
\end{figure}

We now assume that the two mutants present are E255K and F317L.  According to the \emph{in vitro} IC50 value reported in~\cite{redaelli2009activity}, E255K is resistant to each drug. The recommended combination therapy is shown in Table~\ref{tab:2mutants:therapy} below. The combination therapy is different from the combination therapies proposed in the baseline model and the model with M351T and F317L, and is close to the monotherapy with dasatinib.
\begin{table}[htb]
\begin{adjustwidth}{-0.5in}{0in}
\centering
\caption{{\bf The optimal treatment schedules with two mutants.}} 
\begin{tabular}{cc}
\hline
M351T 													& 2 2 2 2 2 2 2 2 2 2 2 2 2 2 2 2 2 2 2 2 2 2 2 2 2 2 2 2 2 2 2 1 1 1 1 1\\
F317L													  & 	2 2 2 2 2 2 2 2 2 1 1 1 1 1 1 1 1 1 1 1 1 1 1 1 1 1 1 1 1 1 1 1 1 1 1 1\\
M351T \& F317L				  & 	2 2 2 2 2 2 2 2 2 1 1 1 1 1 1 1 1 1 1 1 1 1 1 1 1 1 1 1 1 1 1 1 1 1 1 1\\
F317L \& E255K				 &	2 2 2 2 2 2 2 2 2 2 2 2 2 2 2 2 2 2 2 2 2 2 2 2 2 2 2 2 2 2 2 2 2 2 1 1 \\
\hline
\end{tabular}
\label{tab:2mutants:therapy}
\end{adjustwidth}
\end{table}
We also compare the performance of the four different schedules in Fig~\ref{fig:2mutants:therapy}. The leukemic cell population is driven down in the first several months, but increases thereafter due to the increase of E255K population. Since E255K is resistant to each drug, even with the best therapy the leukemic cells still consist of over $73.5\%$ of the total cell population after 36 months.  These results demonstrate that the optimal combination schedule is strongly dependent upon the specific type and combination of preexisting BCR-ABL mutants present at the start of therapy.

\clearpage

\section*{Discussion}
In this work we have considered the problem of finding optimal treatment schedules for the administration of a variety of TKIs for treating chronic phase CML. We modeled the evolution of wild-type and mutant leukemic cell populations with a system of ordinary differential equations, and incorporated a dynamics model of patient ANC level to account for toxicity constraints.
We then formulated an optimization problem to find the sequence of TKIs that lead to a minimal cancerous cell population at the end of a fixed time horizon of 36 months.  The 36 month therapeutic horizon is clinically meaningful since it appears that the risk of therapeutic failure and disease progression to blast crisis is highest within the first two years from diagnosis~\cite{druker2006five}.

At first glance the optimization problem studied in this work (OTP) is quite challenging. It is a mixed-integer nonlinear optimization problem, where the nonlinear constraints are specified by the solution to a nonlinear system of differential equations. However, one factor mitigating the complexity of the problem is the assumption that the TKIs do not effect the stem cell compartment. This has the effect of making the evolution of the stem cell compartment independent of the TKI schedule chosen. In addition, the remaining layers in the cellular hierarchy are modeled by linear differential equations. We can thus numerically solve the differential equation governing the stem cell layer, and treat this function as an inhomogeneous forcing term in the linear differential equation governing the progenitor cells. This allows us to approximate the nonlinear constraints specified by the differential equations by linear constraints with high accuracy. Then the OTP problem can be approximated by a mixed-integer linear optimization problem, which we are able to solve with state-of-the-art optimization software CPLEX~\cite{cplex126} within one hour.

{\it Importance of minimizing progenitor cell population.} We first aimed to minimize leukemic cell burden at 36 months after initiation of therapy, starting with an initial leukemic population of wild-type CML cells and either M351T (sensitive to all three therapies) or F317L (resistant to dasatinib) mutant leukemic cells.  For both starting mutant populations, we observed that the optimal schedule involves initiating therapy with dasatinib and later switching to nilotinib, although the timing of the switch differed.  To further understand this result, we noted that within this parameter regime, dasatinib is the most effective of the three TKIs at controlling leukemic progenitor cells, while nilotinib is the most effective at controlling the differentiated cells, which comprise most of the total leukemic burden. Thus, we note that controlling the leukemic progenitor cell population is important in long-term treatment outcome. This is further supported by the observation that blast crisis emerges due the acquisition of additional mutations in CML progenitor cells (not stem cells)~\cite{jamieson2004granulocyte}.  Our approach suggests that using combination TKI therapies may be a viable method of controlling this population.   Our modeling suggests that it is best to reduce the progenitor cells early and then reduce the differentiated cells towards the end of the treatment planning horizon. An early reduction in progenitor cells pays off in later stages of the treatment planning horizon, since a small progenitor cell population results in a lower growth rate for differentiated cells which leads to a greater response to subsequent TKI therapy.

{\it Effects of toxicity constraints.}
We also imposed a toxicity constraint on therapy optimization procedure by mandating that patient ANC levels stay above a given threshold that reduces the risk of infections. We observed that incorporating this toxicity constraint does impact the structure of the optimal schedules significantly, resulting in mandated treatment breaks as well as switching some months to imatinib therapy, which has a lower toxicity effect.  We also noted that the choice of treatment breaks occurring also in one-month intervals may result in dangerous rebound of leukemic burden to levels close to pre-treatment, suggesting that shorter breaks to combat toxicity may be recommended.  Although the model we have used for describing the dynamics of the ANC levels is simple, our findings demonstrate that incorporating a mechanistically modeled toxicity constraint into optimization of therapy scheduling is both feasible and important in determining optimal scheduling.  

{\it Multiple preexisting mutant types.}  While some previous studies have suggested that the majority of CML patients are diagnosed with 0 or 1 preexisting BCR-ABL mutations, some patients do harbor multiple mutants at the start of therapy \cite{iqbal2013sensitive, leder2011fitness}.  Thus we also investigated the impact of having 2 mutant types present (M351T and F317, or E255K and F317L) at the start of therapy, on optimal schedules.  We observed the number and specific combination of preexisting mutants present can significantly impact the optimization results.  This points to the importance of determining which BCR-ABL mutations preexist in patients at diagnosis, before treatment planning is done.

\vspace{0.1in}


Throughout this work we have observed that the structure of the optimal therapy depends heavily on model parameters, e.g., cellular growth rates and ANC decay rates. It is  likely that each individual patients will have unique model parameters, and therefore a unique best schedule. An exciting application of this work would be the development of personalized optimal therapeutic schedules. Determination of (i) the mutant types (if any) present in a patient's leukemic cell population, (ii) growth kinetics of their leukemic cell populations, and (iii) patient ANC level responses  under various TKIs, would enable our optimization framework to build treatment schedules in a patient-specific setting.

\begin{appendices}
\section{Parameters}
\label{sec:PARAM}
In this section we describe the model parametrization for the examples shown above. A major source of our parameters is the work \cite{olshen2014dynamics} which statistically fit a hierarchical differential equation model (similar to \eqref{eq:hierarchical}) to time series data of CML patents undergoing TKI therapy. 

\subsection{Stem cell kinetics} \label{appendix:sc}
\begin{itemize}
	\item Density dependence parameters $\phi_i$ of type $i$ stem cell, for each $i$. We have $\phi_i =1/(1+p_i\sum_{i=1}^n x_{1,i}(t))$, with $p_1=(b^0_{1,1}/d^0_{1,1}-1)/K_1$, $p_2 = (b^0_{1,2}/d^0_{1,2}-1)/K_2$, and $p_i=p_2$ for $i\ge 3$. The values of $K_1$ and $K_2$ are given in Section~\ref{appendix:initial}.
	\item The birth rates $b^j_{1,i}$. The estimates $b^j_{1,1}=0.008$ and $b^j_{1,i}=0.01$ for any cell type $i\ge 2$ and drug $j$. The value 0.01 is used in~\cite{olshen2014dynamics} for the birth rate of leukemic stem cells without drug. We further assume that this value remains the same under any therapy, which is different from~\cite{olshen2014dynamics}. 
	\item The death rates $d^j_{1,i}$. The estimate $d^j_{1,i}=0.0005$ for any $i$ and $j$, from~\cite{olshen2014dynamics}.
\end{itemize}

\subsection{Progenitor cell kinetics}
\begin{itemize}
		\item The death rates $d^j_{2,i}$. The estimates $d^1_{2,i}=0.0028$ and $d^2_{2,i}=0.0053$ for any $i$, from~\cite{olshen2014dynamics}. The death rate of leukemic progenitor cells under high-dose imatinib ($800$ mg/day) is 0.0035 in~\cite{olshen2014dynamics}. We consider imatinib with regular dose ($400$ mg/day) in this paper, so we set $d^3_{2,i}=0.0035/2=0.00175$ for any $i$. Note the death rates are the same across all cell types with the same therapy, but vary with different therapies. In addition, we set the death rate of normal progenitor cells $d^0_{2,i}=\min\{d^1_{2,i}, d^2_{2,i},d^3_{2,i}\}=0.00175$ for any $i$.
		\item The differentiation rates $b^j_{2,i}$. 
		\begin{itemize}
			\item  For normal cells, $b^j_{2,1}=0.35$ for any $j$. 
			\item For wild type, $b^0_{2,2}=2b^j_{2,1} = 0.70$, $b^1_{2,2}=b^0_{2,2}/400=0.00175$, $b^2_{2,2}=b^0_{2,2}/200=0.0035$, and $b^3_{2,2}=b^0_{2,2}/400=0.00175$. All estimates are from~\cite{olshen2014dynamics}.	
			\item For mutants, the differentiation rates are listed in Table \ref{table:mutant:rate}. Since there are little \emph{in vivo} data available in the literature related to leukemic mutant birth rate, our estimation is based on \emph{in vitro} data for these mutants, in particular the IC50 values. We use a piecewise linear interpolation to estimate the differentiation rates, based on the relative IC50 values of mutants under nilotinib, dasatinib, and imatinib reported in~\cite{redaelli2009activity}. For sensitive or moderately resistant mutants (the relative IC50 value is less than or equal to 4), the differentiation rate of mutant $i$ is estimated using the linear interpolation 
			\[b^j_{2,i}=\text{relative IC50 value of mutant $i$ under drug $j$} \times b^j_{2,2}.\]
For resistant mutants (the relative IC50 value is between 4.01 and 10), the differentiation rate is estimated with the following linear interpolation:
			\begin{equation*} 
			b^j_{2,i} = 0.9 b^{4}_{2,2} + \frac{0.1 b^4_{2.2}}{10-4.01} (\text{relative IC50 value of mutant $i$ under drug $j$} - 4.01).			
			\end{equation*}
Thus if the relative IC50 value for a resistant mutant is 4.01, then its differentiation rate is $90\%$ of the differentiation rate of the WT cell without any drug (0.7 per day); if the relative IC50 value for a resistant mutant is 10, then its differentiation rate is equal to the birth rate of the WT progenitor cell without drug. For highly resistant mutants (the relative IC50 is larger than 10), we set its differentiation rate to the differentiation rate of the WT progenitor cells without drug. 			
			\begin{table}[htb]
			\centering
			 \caption{The differentiation rate of mutant progenitor cells under three drugs} \label{table:mutant:rate}
			 \begin{tabular}{ccccccc}
			 \hline
			   & E255K & E255V & F317L & M351T & Y253F & V299L\\
			 \hline
			 Nilotinib ($b^1_{2,i}$) & 0.6614	 & 0.7	& 0.00389	& 0.00077	&0.00565 & 0.00235\\
			 Dasatinib ($b^2_{2,i}$) & 0.6488 & 0.0120	& 0.6354   &	0.00308	& 0.00553 & 0.6843\\
			 Imatinib ($b^3_{2,i}$)  & 0.6536  &0.7   &	0.00455	& 0.00308	& 0.00627 & 0.00270\\
			 \hline
			 \end{tabular}
			\end{table}
		\end{itemize}			
\end{itemize}		

\subsection{Differentiated cell kinetics}
\begin{itemize}	
		\item The death rate $d^j_{3,i}$. The estimates $d^1_{3,i}=0.0442$ and $d^2_{3,i}=0.0394$ for any $i$, from~\cite{olshen2014dynamics}. The death rate of leukemic differentiation cells under high-dose imatinib ($800$ mg/day) is $0.055$ in~\cite{olshen2014dynamics}. We consider imatinib with regular dose ($400$ mg/day) in this paper, so we set $d^3_{3,i}=0.055/2=0.0275$ for any $i$. In addition, $d^0_{3,i}=\min\{d^1_{3,i}, d^2_{3,i},d^3_{3,i}\}=0.0275$ for any $i$.		
		\item The differentiation rates $b^j_{3,i}$.
		\begin{itemize}
			\item For normal cells, $b^j_{3,1}=5.5$ for any $j$. 
			\item For wild type, $b^0_{3,2}=1.5b^0_{3,1} = 8.25$, $b^1_{3,2}=b^0_{3,2}/600=0.01375$, $b^2_{3,2}=b^0_{3,2}/300=0.0275$, and $b^3_{3,2}=b^0_{3,2}/600=0.01375$. All estimates are from~\cite{olshen2014dynamics}.	
			\item For the mutant, if it is sensitive or moderately resistant to drug $j$ (the relative IC50 value is less than or equal to 4), then $b^j_{3,3}= b^j_{2,3} \times b^j_{3,2}/b^j_{2,2}$, for $j=1,2,3$; otherwise $b^j_{3,3}= b^j_{2,3} \times b^4_{3,2}/b^4_{2,2}$, for $j=1,2,3$.
		\end{itemize}
\end{itemize}

\subsection{Terminally differentiated cell kinetics} \label{appendix:tc}
Using the estimates from~\cite{olshen2014dynamics}, we set the differentiation rates $b^j_{4,i}=100$ and death rates $d^j_{4,i}=1$ for any $i$ and $j$.

\subsection{Initial cell populations at diagnosis} \label{appendix:initial}
The normal marrow output in an adult is approximately $3.5\times 10^{11}$ cells per day~\cite{dingli2008chronic}. To achieve this equilibrium condition, we set $K_1 =8.75 \times 10^4$ in differential equations~\eqref{eq:hierarchical} with parameters described in Sections~\ref{appendix:sc} to~\ref{appendix:tc} and in the absence of leukemic cells. To obtain an estimate of $K_2$, we assume that diagnosis of CML occurs once the leukemic cell burden reaches a threshold of $10^{12}$ cells~\cite{holyoake2002elucidating}, and that the differential equations~\eqref{eq:hierarchical} have parameters described in Sections~\ref{appendix:sc} to~\ref{appendix:tc} and start with $K_1=8.75\times 10^4$ normal stem cells, one wild-type leukemic stem cell, and no other cells. We set $K_2=3\times 10^6$ so that the patient is diagnosed with CML around 78 months (6.5 years) after the first leukemic stem cell arises. At diagnosis, the normal stem cell, progenitor cell, differentiated cell, and terminally differentiated cell populations are $7.34\times 10^4$, $1.61\times 10^7$, $3.24\times 10^9$, and $3.24\times 10^{11}$, respectively; the leukemic stem cell, progenitor cell, differentiated cell, and terminally differentiated cell populations are $2.95\times 10^5$, $4.07\times 10^7$, $1.08\times 10^{10}$, and $1.08\times 10^{12}$ respectively. These are used as the initial cell populations for a patient diagnosed with CML.

\subsection{ANC kinetics}
	\begin{itemize}
		\item We require the patient's ANC cannot fall below $\Le=1000/\mathrm{mm}^3$, the normal level of ANC is $\Ue=3000/\mathrm{mm}^3$, and the patient's initial ANC is $3000/\mathrm{mm}^3$.
		\item It is observed that nilotinib has higher toxicity than imatinib~\cite{cortes2010nilotinib}. We set the estimated monthly decrease rates of ANC to be $d_{anc,1}=350/\mathrm{mm}^3$ under nilotinib, $d_{anc,2}=300/\mathrm{mm}^3$ under dasatinib, and $d_{anc,3}=250/\mathrm{mm}^3$ under imatinib. The ANC of a patient increases by $b_{anc} = 2000/\mathrm{mm}^3$ during a drug holiday, before it reaches the normal level $3000/\mathrm{mm}^3$. We also investigate how optimal schedule is affected if dasatinib has a higher toxicity than nilotinib, with $d_{anc,1}=300/\mathrm{mm}^3$ and $d_{anc,2}=350/\mathrm{mm}^3$.
	\end{itemize}

\section{Method to solve the optimization model}
\label{sec:MILO}
We describe the method to solve the optimization model~\eqref{eq:OTP} introduced in Section~\ref{sec:optmodel}. Our strategy is to build a mixed-integer linear optimization model~\cite{nemhauser1999integer} that approximates the optimization model~\eqref{eq:OTP}, and then solve the approximation model to optimality numerically by off-the-shelf optimization software CPLEX~\cite{cplex126}. The mixed-integer linear optimization model is built through two steps: (1) we first approximate the ODE constraints~\eqref{eq:OTP:ode} by bilinear constraints; (2) we then transform the bilinear constraints and nonlinear constraints~\eqref{eq:OTP:tox1} into equivalent linear constraints, by adding auxiliary decision variables.

We first describe how to approximate the ODE constraints~\eqref{eq:OTP:ode} by bilinear constraints.  
Suppose patients take drug $j$ in month $m$. Since the cell dynamics are modeled by the following set of ODEs
\begin{subequations}  \label{eq:odeconcise}
\begin{align}
\dot{x}(t)=f^j(x(t)), \ & t\in [m\deltat, (m+1)\deltat],\\
x(m\deltat) = x^m, \ &
\end{align}
\end{subequations}
the cell abundances in month $m+1$, $x^{m+1}$, are completely determined by the initial cell abundance $x^m$ and function $f^j$. Without loss of generality, we assume this relationship is described by
\begin{equation}\label{eq:odesol}
x^{m+1}_{l,i}=g^j_{l,i}(x^{m})
\end{equation}
with some unknown nonlinear function $g^j_{l,i}:\Re^{Ln} \rightarrow \Re$, for each month $m$, layer $l$, and cell type $i$. Recall that $L$ is total number of cell layers ($L=4$), and $n$ is the total number of cell types. Then the ODE constraints~\eqref{eq:OTP:ode} are equivalent to the constraints below
\begin{equation} \label{eq:OTP:odenlp}
x^{m+1}_{l,i} = \sum_{j\in \J} z^{m,j}g^j_{l,i}(x^{m}), \text{ for each } m, l, i.
\end{equation}
We will approximate the nonlinear function $g^j_{l,i}$ with an affine function $\hat{g}^{m,j}_{l,i}: \Re^{Ln} \rightarrow \Re$, for each $m,j,l$, and $i$. In particular, the function 
\begin{equation} \label{eq:ode:affine}
\hat{g}^{m,j}_{l,i}(x)=a^{j,l,i}x+h^{m,j}_{l,i},
\end{equation}
where $a^{j,l,i}$ is an $(Ln)$-dimensional vector and does not depend on $m$. Details of how $\hat{g}^{m,j}_{l,i}$ is constructed are provided in Section~\ref{sec:LIN_ODE} of the Appendix. Let $a^{j,l,i}=[a^{j,l,i}_{1,1}, \ldots, a^{j,l,i}_{k,s}, \ldots, a^{j,l,i}_{L,n}]$. Then constraint~\eqref{eq:OTP:odenlp} can be approximated by the bilinear constraint
\begin{equation}\label{eq:OTP:bilinear}
 x^{m+1}_{l,i} = \sum_{j\in \J} z^{m,j}\hat{g}^{m,j}_{l,i}(x^{m}) =\sum_{j\in \J} z^{m,j}(\sum_{k\in \L, s \in \I} a^{j,l,i}_{k,s} x^m_{k,s}+h^{m,j}_{l,i}),
\end{equation}
for each type $i$ cell at layer $l$ in month $m$. 

We now describe how to transform bilinear constraints~\eqref{eq:OTP:bilinear} and piecewise linear constraints~\eqref{eq:OTP:tox1} into linear constraints. These are standard techniques in mixed-integer linear optimization~\cite{nemhauser1999integer}. We introduce auxiliary continuous variables $v^{m,j}_{k,s}$, and set $v^{m,j}_{k,s}=z^{m,j}x^m_{k,s}$. Then bilinear constraints~\eqref{eq:OTP:bilinear} are transformed into the equivalent linear constraints below.
\begin{equation} \label{eq:OTP:odelin}
\begin{split}
x^{m+1}_{l,i} =& \sum_{j\in \J} (\sum_{k\in \L, s \in \I} a^{j,l,i}_{k,s} v^{m,j}_{k,s}+h^{m,j}_{l,i}z^{m,j})\\
0 \le &v^{m,j}_{k,s} \le U_{k,s}z^{m,j}, \\
0 \le &  x^m_{k,s}  - v^{m,j}_{k,s}  \le U_{k,s}(1-z^{m,j}),
\end{split}
\end{equation}
where $U_{k,s}$ is an upper bounds of cell abundance $x^m_{k,s}$ for each $m$. The value of $U_{k,s}$ can be obtained by taking the maximum value of layer $k$ type $s$ cell abundances over the whole planning horizon under all three monotherapies and no treatment. The piecewise linear constraints~\eqref{eq:OTP:tox1} can be transformed into equivalent linear constraints below, by introducing auxiliary continuous variable $u^m$ and binary variable $q^m$ for each $m$.
\begin{subequations} \label{eq:OTP:toxlin}
\begin{align}
u^{m+1} = y^{m} + \b - \sum_{j\in \J \setminus\{0\}} \dj z^{m,j}, \\
y^{m+1} \ge u^{m+1} - b_{anc}q^{m+1},\\
y^{m+1} \ge \Ue - (\Ue - \Le) (1-q^{m+1}), \\
y^{m+1} \le u^{m+1},\\
y^{m+1} \le \Ue,\\
q^{m+1} \in \{0,1\}.
\end{align}
\end{subequations}

\noindent Overall, the optimization model~\eqref{eq:OTP} is approximated by the following mixed-integer linear optimization model.
\begin{subequations} \label{eq:OTP:MILP}
\begin{align}
	\min \ &\sum_{l \ge 1} \sum_{i\ge 2} x^M_{l,i}\\
	\text{s.t.} \ & x^{m+1}_{l,i}= \sum_{j\in \J} \sum_{k\in \L, s \in \I} a^{j,l,i}_{k,s} v^{m,j}_{k,s}+\sum_{j\in \J}h^{m,j}_{l,i}z^{m,j}, &i\in \I, l\in \L, m \in\m  \\
& 0 \le v^{m,j}_{k,s} \le U_{k,s}z^{m,j}, &i\in \I, l\in \L, m \in\m\\
& 0 \le  x^m_{k,s}  - v^{m,j}_{k,s}  \le U_{k,s}(1-z^{m,j}), &i\in \I, l\in \L, m \in\m\\
&u^{m+1} = y^{m} + \b - \sum_{j\in \J \setminus\{0\}} \dj z^{m,j}, & m\in \m\\
&y^{m+1} \ge u^{m+1} - b_{anc}q^{m+1},  & m \in \m\\
&y^{m+1} \ge \Ue - (\Ue - \Le) (1-q^{m+1}), & m\in \m \\
&y^{m+1} \le u^{m+1}, & m\in \m\\
&y^{m+1} \le \Ue, & m\in \m\\
& y^m \ge \Le, & m\in \M\\
& \sum_{j\in \J} z^{m,j} = 1,  &m\in \m \\ 		
& z^{m,j}, q^{m+1}\in \{0,1\}, & m\in \m, j\in \J\\ 
& x(0) = x^0, \; y^0  \text{ is given}.
\end{align}
\end{subequations}

\section{Linear approximation to the solutions of the ODEs}
\label{sec:LIN_ODE}
In this section, we describe how to construct the affine function $\hat{g}^{m,j}_{l,i}$ in~\eqref{eq:ode:affine} in Section~\ref{sec:MILO} of the Appendix. If we assume that the drugs do not affect stem cells, we can compute the abundance of stem cells over the planning horizon numerically in advance, regardless of the treatment schedules. Thus we assume $x_{1,1}(t), \ldots, x_{1,n}(t)$ are given as data, for any $t$. We can first eliminate all the variables $x^m_{1,i}$ and constraints containing $x^m_{1,i}$, for each $m$ and $i$, in the optimization problem~\eqref{eq:OTP:MILP}. The dynamics of wild-type leukemic cells and each mutant type have no impact on each other. We can decouple the ODEs into a series of linear ODEs as follows, each describing the dynamics for type $i$ cell from layer 2 to layer 4. 
\begin{equation} \label{eq:ode:lineage}
\left[
\begin{array}{c}
\dot{x}_{2,i}(t) \\
\dot{x}_{3,i}(t) \\
\dot{x}_{4,i}(t)\\
\end{array}
\right]
=
\left[
\begin{array}{ccc}
-d^j_{2,i} & 0 & 0 \\
b^j_{3,i} & -d^j_{3,i} & 0\\
0 & b^j_{4,i} & -d^j_{4,i}\\
\end{array}
\right]
\left[
\begin{array}{c}
x_{2,i}(t) \\
x_{3,i}(t) \\
x_{4,i}(t)\\
\end{array}
\right]
+
\left[
\begin{array}{c}
b^j_{2,i} x_{1,i}(t) \\
0 \\
0\\
\end{array}
\right]
\end{equation}

\noindent Write the above equations in the matrix form, we have
\begin{equation} \label{eq:ode:matrix}
\dot{v}_i(t) = W^j_i v_i(t) + w^j_i(t), \text{ for } t\in [m\deltat, (m+1)\deltat],
\end{equation}
where $v_{i}(t)=[x_{2,i}(t), x_{3,i}(t), x_{4,i}(t)]^{\top}$, $w^j_i(t)=[b^j_{2,i}x_{1,i}(t), 0,0]^{\top}$, and $W^j_i$ is the lower triangular matrix in~\eqref{eq:ode:lineage}. 

We divide $(m\deltat, (m+1)\deltat)$ into $\deltat=30$ one-day sub-intervals. Consider a sub-interval $(t_0, t_0+1)$. By assuming $w^j_i(t) = w^j_i(t_0)$ for any $t\in (t_0, t_0+1)$, we solve~\eqref{eq:ode:matrix} approximately and obtain
\begin{equation} ~\label{eq:ode:oneday}
v_i(t_0+1) \approx e^{W^j_i}v_i(t_0) + (e^{W^j_i} - I)(W^j_i)^{-1} w^j_i(t_0).
\end{equation}
By combining equations~\eqref{eq:ode:oneday} for $t_0=m\deltat, m\deltat+1,\ldots, (m+1)\deltat-1$, we have
\begin{equation} \label{eq:linapprox}
v_i((m+1)\deltat) = e^{W^j_i \deltat}v_i(m \deltat) + \sum_{d=0}^{\deltat-1}e^{W^j_i(\deltat-1-d)}(e^{W^j_i} - I)(W^j_i)^{-1} w^j_i(m \deltat+d).
\end{equation}

\noindent Recall that $v_{i}(m\deltat)=[x^{m}_{2,i}, x^{m}_{3,i}, x^m_{4,i}]^{\top}$ for each $m$. Thus~\eqref{eq:linapprox} can be rewritten as
\begin{equation} \label{eq:linapprox:concise}
\left[
\begin{array}{c}
x^{m+1}_{2,i}\\
x^{m+1}_{3,i} \\
x^{m+1}_{4,i} 
\end{array}
\right]
 = A^{j,i}
\left[
\begin{array}{c}
x^{m}_{2,i}\\
x^{m}_{3,i} \\
x^{m}_{4,i} 
\end{array}
\right]
+ h^{m,j}_i,
\end{equation}
where $A^{j,i}=(e^{W^j_i})^{\deltat}$ and $h^{m,j}_i= \sum_{d=0}^{\deltat-1}(e^{W^j_i})^{\deltat-1-d}(e^{W^j_i} - I)(W^j_i)^{-1} w^j_i(m\deltat+d)$. Each equation in~\eqref{eq:linapprox:concise} is used as the affine function $\hat{g}^{m,j}_{l,i}$ in~\eqref{eq:ode:affine}, for each $m,j,i$, and $l=2,3,4$.
\end{appendices}

%
%
%

\end{document}